**Biological Regulatory Network Inference through Circular Causal Structure Learning**

Hongyang Jiang[1], Yuezhu Wang[2], Ke Feng[2], Chaoyi Yin[2], Yi Chang[2,3,4,*], Huiyan Sun[2,3,4,*]

1 Data Science & Artificial Intelligence Research Institute, China Unicom

2 School of Artificial Intelligence, Jilin University, Changchun 130012, China

3 International Center of Future Science, Jilin University, Changchun, China,

4 Engineering Research Center of Knowledge-Driven Human-Machine Intelligence, MOE, China

* Correspondence: huiyansun@jlu.edu.cn (H.S.);

## Abstract

Biological networks are pivotal in deciphering the complexity and functionality of biological systems. Causal inference, which focuses on determining the directionality and strength of interactions between variables rather than merely relying on correlations, is considered a logical approach for inferring biological networks. Existing methods for causal structure inference typically assume that causal relationships between variables can be represented by directed acyclic graphs (DAGs). However, this assumption is at odds with the reality of widespread feedback loops in biological systems, making these methods unsuitable for direct use in biological network inference. In this study, we propose a new framework named SCALD (Structural CAusal model for Loop Diagram), which employs a nonlinear structure equation model and a stable feedback loop conditional constraint through continuous optimization to infer causal regulatory relationships under feedback loops. We observe that SCALD outperforms state-of-the-art methods in inferring both transcriptional regulatory networks and signaling transduction networks. SCALD has irreplaceable advantages in identifying feedback regulation. Through transcription factor (TF) perturbation data analysis, we further validate the accuracy and sensitivity of SCALD. Additionally, SCALD facilitates the discovery of previously unknown regulatory relationships, which we have subsequently confirmed through ChIP-seq data analysis. Furthermore, by utilizing SCALD, we infer the key driver genes that facilitate the transformation from colon inflammation to cancer by examining the dynamic changes within regulatory networks during the process.

## Introduction

Biological networks serve as effective tools for modeling the intricate regulatory relationships among biomolecules, such as gene regulatory networks (GRNs), signal transduction networks (STNs), protein-protein interaction networks (PPIs), metabolic networks, etc [1]. Taking GRN inference as an example, most of the existing network construction methods are based on generalized correlation and are roughly divided into the following categories: (i) regression-based methods (TIGRESS [2], Genie3 [3], GRNBOOST2 [4], and PLSNET [5]); (ii) mutual information-based methods (CLR[6], ARACNE[7], and PIDC[8]); (iii) Bayesian network methods (catnet[9], MMPC [10], and BMA[11]); (iv) ordinary differential equation (ODE)-based methods (NARROMI[12] and CNNC[13]); and (v) deep learning-based methods (3DCEMA [14] and DeepSEM[15]). However, in biological system, regulation is essentially a causal relationship with directionality, which is not enough to describe by correlation alone. In addition, generalized correlation-based methods usually lead to a large number of redundant and spurious edges due to the existence of confounders. Therefore, it is more reasonable to infer GRN from the perspective of causality to establish the regulatory relationship[16].

Although time series data often facilitates causal inference by leveraging the principle that the cause precedes effect, acquiring high-quality time series biological data poses a significant challenge. Therefore, causal discovery based on non-time series observational data becomes a more universally applicable approach. Causal discovery approaches utilizing non-time series observational data typically assume that the causal graph adheres to a Directed Acyclic Graph (DAG) structure, and they primarily fall into three categories: constraint-based methods[17]–[19], causal function-based methods[20]–[22], and score-based methods[23]–[28]. Constraint-based methods infer causality through conditional independent tests between variables, under the Causal Markov Assumption and the Causal Faithfulness Assumption. Commonly used methods in this category include PC[17], IC[18], and FCI[19]. However, the presence of Markov equivalence classes always makes causal directionalities unidentifiable. Causal function-based methods, proposed on the basis of Structural Equation Model (SEM)[35] and the Causal Sufficiency Assumption, orient edges by analyzing the asymmetry in regression residual between variables, as seen in models like Additive Noise Model (ANM) [21] and post-nonlinear model (PNL)[22]. Such an approach effectively addresses the issue of unidentifiability associated with Markov equivalence class. Score-based approaches, like Greedy Equivalence Search (GES)[23], evaluate potential causal graphs using various scoring methods, including the Bayesian Information Criterion (BIC) [24] and Bayesian Gaussian Equivalence (BGe)[25]. The graph with the highest score is ultimately selected. However, these above methods obtaining DAGs essentially involve a combinatorial search. As the number of variables increases, the complexity of this search escalates exponentially, posing significant challenges when attempting to find optimal solutions for large-scale DAGs. In contrast to traditional methods that typically utilize localized heuristics to maintain acyclicity, the continuously optimizable DAG constraint is able to transform the search process into a continuously optimizable equality constraint, such as Notears[27] and DAG-GNN[28] which uncovers nonlinear causal relationships by employs a Variational Autoencoder (VAE) model.

In recent years, causal discovery technology has been gradually applied to biological network inference. Inspired by DAG-GNN, DeepSEM[15] applies variational autoencoders to learn low-dimensional representations of genes and deploys structural equation model to predict the strength of regulatory relationships. Zhang et al.[29] proposes a causal discovery method based on PC algorithm, DirectTarget, specifically designed to infer miRNA-mRNA causal regulatory relationships in heterogeneous data. Although existing causal discovery methods under the DAG hypothesis have proven effective in inferring regulatory networks, they ignore the circular structures formed by numerous feedback loops in biological systems. Moreover, maintaining regulatory system stability is critical, as failure to do so could lead to system expansion or functional contraction. Hence, when inferring biological networks from a causality perspective, it's essential to take into account the feedback mechanism and its role in stabilizing regulation. In light of this, a few methods have been proposed to explore the inference of Directed Cyclic Graph (DCG) [30], [31]. Bongers has extended the acyclic SEM(Structural Equation Model) to a cyclic SEM and provided a comprehensive set of theorems for DCG-based structural equation models[31]. This work demonstrates that, with the exception of self-loops, all other loops in the DCG can be represented by causal structural models. However, it's worth noting that the stable regulation of feedback mechanisms in biological systems is seldom addressed in existing studies of causal regulatory network inference.

In this study, we propose SCALD (Structural CAusal model for Loop Diagram), a novel framework for inferring biological network causal structures under feedback regulation. This approach synergistically combines a nonlinear structure equation model, which represents the relationships between variables, with stable feedback control conditions through continuous optimization. To our knowledge, this is the first study that integrates causal structure discovery with feedback loop considerations to infer directed biological networks. We conduct extensive experiments and comparisons with existing benchmark methods to evaluate SCALD's performance in identifying regulatory relationships between molecules for both gene regulatory networks and signaling transduction networks. Further, through public transcription factor perturbation experiments, we assess the sensitivity and accuracy of SCALD in predicting transcriptional regulatory relationships between transcription factors and their corresponding target genes. For the newly predicted unknown regulatory relationships, we provide additional validation through ChIP-seq data analysis. Beyond validation, we also explore downstream applications of SCALD. For instance, we investigate its potential in identifying key drivers that promote the transformation from colon inflammation to cancer by recognizing dynamic changes in regulatory networks during the process. The data and code are available at https://github.com/JhyOnya/SCALD.

## Results

### Overview of SCALD

The SCALD framework is composed of two main components: (i) a nonlinear circular structural equation model, based on neural networks, which is designed to extract the foundational structure of a causal regulatory network; and (ii) network topology constraints, which are formulated to characterize the causal relationships of regulation and aim to eliminate unstable loops in the presence of feedback loops. To tackle the optimization challenges inherent in this framework, we utilize an augmented Lagrangian method. In terms of the complexities of biological networks, where causal relationships are inherently nonlinear and directed, we integrate a two-layer fully connected neural network with the Exponential Linear Unit (ELU) activation function for the nonlinear structural equation model. As most current methods for causal discovery are limited by the assumption of directed acyclic graphs (DAGs), an assumption that often conflicts with the cyclic nature of biological networks. we delve into the topological features of stable feedback loops, incorporating symbolic information that denotes either promotional or inhibitory interactions. Consequently, we introduce constraints that adapt our causal network structure inference framework to a more realistic scenario that includes feedback loops. The overall architecture of the SCALD framework is illustrated in Fig. 1.

## SCALD performs well in inferring both gene regulatory network and signaling network from bulk and single cell data

To evaluate SCALD's effectiveness in GRN inference, we first conducte experiments on DREAM5 datasets for S.cerevisiae and E.coli. We then compare SCALD with 11 widely-used GRN methods and one causal discovery method, which requires fewer sample sizes. These benchmark methods encompass five categories: regression-based methods (TIGRESS[2], Genie3[3], GRNBOOST2[4], and PLSNET[5]), mutual information-based method (CLR[6]), Bayesian network method (BMA[11]), ODE-based methods (NARROMI[12]), correlation-based methods (pearson[32], pairwise entropy, conditional entropy, and PPCOR[33]), along with a causal discovery algorithm (ANM). To measure inference performance, we utilized three evaluation metrics: Early Precision Ratio (EPR), Area Under the Precision-Recall Curve (AUPR), and Area Under the Receiver Operating Characteristic Curve (AUROC). In the case of S.cerevisiae data, SCALD outperforms all 11 benchmarks in terms of EPR, 82%( 9 out of 11) benchmarks under AUPR, and all benchmarks in terms of AUROC. When it comes to E.coli data, SCALD surpasses all benchmarks under both EPR and AUPR metrics, and 91% (10 out 11) of benchmarks in terms of AUROC. Following SCALD, GRNBOOST2 and PLSNET ranked second and third under EPR and AUPR, and AUROC, respectively (Fig. 2A, Fig. 2B, and Supplementary Fig. S1).

We further validated the performance of SCALD using single-cell data. Given that most inference methods primarily consider the existence of correlation relationships, we also compare SCALD with existing benchmark causal structure learning methods to assess its ability to accurately determine the directionality of regulatory relationships. Hence, we compared SCALD with six causal discovery methods and one GRN method using hESC scRNA-seq data obtained from the BEELINE benchmark repository. These causal discovery methods include ANM[21], LiNGAM[20], DirectLiNGAM[34], DAG-GNN[28], Notears-linear[27], and Notears-nonlinear[35]. And GRNBoost2 is acknowledged as a highly efficient and stable GRN benchmark method. We create two subsets of genes, each containing either the 500 or 1,000 most variable genes. We then collect three different types of ground-truth network: Cell-type-specific, Nonspecific, and STRING. Specifically, Cell-type-specific and Nonspecific data are derived from Pratapa et al.[36], while STRING data are sourced from functional connectivity within the STRING database[37]. By intersecting each ground-truth network with each subset of genes, we conduct six experiments (CellType-500, NonSpecific-500, STRING-500, CellType-1000, NonSpecific-1000, STRING -1000) for GRN inference analysis. These subsets are named based on ground-truth networks and the number of most significantly changed TGs (500 or 1000). As shown in Fig. 2C, SCALD ranks first in at least two of the three metrics in a data set across these six experiments. In addition, we analyze the functions of the top 50 genes with the highest Degree Centrality nodes in the network constructed by SCALD through GO (Gene Ontology) and KEGG (Kyoto Encyclopedia of Genes and Genomes) enrichment analysis. The results reveal that these genes are significantly enriched in Signaling pathways regulating pluripotency of stem cells, Stem cell differentiation, TGF-β signaling pathway, PI3K-Akt signaling pathway, and Wnt signaling pathway. All these pathways are highly associated with embryonic stem cells[38]–[47], which indicates the capability of SCALD in predicting regulatory association to some extent (Fig. S2 and Table S1).

Additionally, we also evaluate the performance of SCALD in inferring the direction of transmission of biological signals within signaling networks using the Sachs dataset. This is done by comparing it with seven causal discovery methods and one GRN inference method as mentioned above. The Sachs dataset consists of simultaneous measurements of 11 phosphorylated proteins and phospholipids, derived from thousands of individual primary immune system cells under both general and specific molecular interventions. Based on the probability values, we select the top 20 edges (equivalent to the number of edges in the ground truth network) to construct the predicted protein signaling network. The left side of the Fig. 2D displays the ground truth for the protein signaling network of the Sachs dataset, while the right side presents the network predicted by our model. Our algorithm accurately identifies 9 edges (depicted as solid green lines) that correspond to the true connections in the network. Additionally, it detects 5 indirectly connected edges (dashed yellow lines) and 4 reversed edges (solid red lines). Our prediction also included 2 incorrect edges, represented by dashed blue lines. As shown in Fig. 2E, SCALD outperforms the benchmark algorithms. The specific number of predicted edges, along with the evaluation results for EPR, AUPR, and AUROC are provided in the Supplementary Table S2.

## SCALD enables the inference of regulatory relationship with feedback loops

Feedback regulation is a common phenomenon in biological systems. However, causal structure discovery methods, under the Directed Acyclic Graph (DAG) assumption, are incapable of capturing these loops, and generalized correlation-based methods can hardly detect directionality. We evaluate SCALD's ability in identifying feedback loops by comparing its results with the established ground truth.

Initially, we perform a statistical analysis of the loops in the gold standard biological networks within the BEELINE repository. Given the intricacy in determining the exact number of loops in cases with nested loops, we perform a straightforward statistical analysis focusing on the number of nodes within the loops of the gold standard network. This is accomplished through the following steps: we first calculated the connectivity between nodes via adjacent matrix multiplication $\mathbf{D} = (\mathbf{I} + \mathbf{A})^d$ (as described in Methods Section 4.3), and then examined whether each node can point to itself through a path of length d, as denoted by $\gamma_i = \begin{cases} 1, \text{if } \mathbf{D}_{ii} > 1 \\ 0, \text{if } \mathbf{D}_{ii} = 1 \end{cases}$, where $\gamma \in ¡^d$. $\sum \gamma_i$ denotes the number of nodes on the loop. Given that target genes can only be regulated by, and not regulate, transcription factors (TFs) and are thus incapable of forming feedback loops, only TFs have the potential to initiate such regulatory loops. We provide statistical information on nodes within loops, the total number of genes, and the number of isolated genes in the ground truth networks in the BEELINE repository, as detailed in Supplementary Table S3. Our observations reveal that approximately 10% of nodes participate in loops. Additionally, we extract all the TFs and their connections in the loop in the gold standard network for each baseline dataset, assigning symbols to the edges in the loop based on Spearman correlation coefficient. Fig. 3A illustrates examples of partially stable loops in CellType-500, CellType-1000, NonSpecific-500, and NonSpecific-1000, respectively. This indicates that numerous stable loops exist in biological networks, and neglecting these loops can introduce bias in the network inference.

To assess the effectiveness of SCALD in inferring the feedback loops of biological networks, we infer the gene networks constructed by all the TFs extracted from the loops in the gold standard network across six scRNA-seq experiments of human embryonic stem cells. We then compare these inference results with 12 other GRN inference methods (since DAG-based causal structure learning methods are uable to detect loops). Fig. 3B and Fig. 3C display the comparative results on NonSpecific-500 and NonSpecific-1000 experiments, while the results of other four experiments (CellType-500, CellType-1000, STRING-500, and STRING-1000) are shown in Supplementary Fig. S3. These results collectively demonstrate that SCALD outperforms all other benchmark methods in identifying stable loops under the ERP, AUPR, and AUROC metrics.

Moreover, the identification of feedback loops is important for comprehensively understanding the biological system as feedback regulations play pivotal roles in various biological processes. In our analysis of the aforementioned hESC dataset, we have identified multiple regulatory loops and mapped the genes within these circuits to various biological pathways. For example, we find that genes in the "POLR2K-POLR1A-POLR2D-POLR2F-POLR3" directed loop are associated with RNA polymerase, which is a critical component in the transcriptional regulation process of embryonic stem cells[48]. Additionally, we discover that the identified POLR2D-GTF2H2-POLR2F circuit is linked to nucleotide excision repair, which is consistent with existing finding that nucleotide excision repair capacity increases during the differentiation of embryonic carcinoma stem cells [49]. In summary, our research suggest that regulatory feedbacks occupy a significant portion of biological systems and play a vital role in maintaining cell state and essential biological functions. Our approach bridges a gap in the inferring directed biological regulatory networks with feedback loop.

To further validate the efficacy of the loop constraint component on infer biological network, we conduct ablation studies on SCALD, both with and without the loop constraint components, across these six experiments. The observed decline in performance for each individual highlights the need of stable loop constraint module of SCALD. For example, in the CellType-1000 experiment, SCALD with the loop constraint components outperformed the version without them by 38.9% in terms of performance improvement. This resulted in an increase of 9.9% in AUPR and 12.5% in AURPC, as detailed in Supplementary Table S4.

## Validation of SCALD's efficacy in capturing regulatory relationships through transcription factor perturbation data.

Regarding the constructed GRN on gene expression data using SCALD, we further employ perturbation data to validate the trustworthiness of regulatory relationships within biology networks.

For the regulatory relationship predicted by SCALD, we examine whether the expression of a target gene (TG) regulated by a transcription factor (TF) undergoes significant change when TF is knock down or knock out. This examination is conducted using an independent perturbations dataset from the same cell type. We apply SCALD to infer the GRN of CD4+ T cell using the GSE131407 dataset, and utilized perturbed data GSE46333 for validation. As an illustrative example, we focused on the transcription factors ELK1 and NFATC3 in CD4+ T cell.

ELK1, a transcription factor that binds to purine-rich DNA sequences, acts autonomously within the thymus to control the generation of innate-like T cells with memory-like characteristics[50]. Through SCALD, we identify the top 10 candidate target genes of ELK1 in terms of regulatory significance，including *DNAJC1, HOXA4, TRIM56, UBE2F, NR2C2, ETHE1, MIR663B, C4orf41, CREBL2,* and *LGALS8*. Among these, five gene, that is *DNAJC1, HOXA4, TRIM56, ETHE1*, and *LGALS8*, exhibit significantly differential expression (p-value < 0.05 and absolute log2 fold-change > 1) in the data before and after ELK1 knockdown. Several of these genes are strongly associated with immunophenotyping and immunoregulation[51], [52]. For example, *DNAJC1* is known to influence the phenotype of helper T cells, and *LGALS8* plays an important role in tumor immunosuppression, with its high expression correlating with suppressive immune cell expansion in the tumor immune microenvironment[53], [54]. In contrast, for the GRN predicted by GRNBOOST2, only one gene *KCNQ2* out ten genes' expression significantly changes when ELK1 is perturbated.

Another example is transcription factor NFATC3, a member of NFAT family associated with promoting T cell activation and development[55]. *PRMT7* and *SAMHD1*, ranking second and eighth respectively among NFATC3's target genes under SCALD's predictions, exhibit significantly differential expression when NFATC3 is perturbated. Exiting studies have reported that *PRMT7* plays a crucial role in immunomodulation, immune tolerance, and inflammation in regulatory T cells (Tregs)[56]. *SAMHD1*'s expression is reduced during CD4 T cell activation and proliferation, and its reduced expression level predisposes CD4+ T cells to be more sensitive to HIV. [57]. These finding demonstrate the efficacy of SCALD in predicting regulatory relationships.

In our statistical analysis, we count the number of genes that have been predicted as the target genes of certain TF by SCALD and simultaneously show differential expression after the perturbation of these TFs. Fig. 4A presents the result of top 10 and 20 predicted target genes by SCALD and GRNBOOST2 for ELK1 and NFATC3. The results demonstrate that our method is able to identify target genes related to the TFs with greater precision.

### SCALD enable predict unknown regulatory relationships.

In the context of regulatory relationships predicted by SCALD but not present in the established ground truth, we extend our investigation by conducting Chromatin Immunoprecipitation Sequencing (ChIP-seq) data analyses, which allows us to validate the existence of such regulatory relationships. To maintain cell type consistency, we collect ChIP-seq data for transcription factors from the H1 embryonic stem cell line, as provided by ENCODE[58]. By examining the overlapping binding sites of transcription factors (TF) and the promoter regions (TSS -2000~+500) of their target genes as predicted by SCALD, we assess SCALD's capability to predict new TF-target gene (TG) pairs. As a case in point, we concentrate on the EGR1-FLI1 regulatory pair in H1 embryonic stem cells.

Our predictions (as shown in Fig. 5A) reveal that FLI1 is the top-ranked target gene of the transcription factor EGR1, with an exceptionally significant p-value throughout the inferred network. Notably, this pair is absent in the current version of the ground truth GRN. Through conducting a motif scan analysis on the genomic sequence of the FLI1 promoter region, we find a significant EGR1 motif within the binding site region (chr11:128686332-128686592) located upstream of the FLI1 transcription start site (chr11:128684535-128687034) with p value=1.24E-5 (in Fig. 5B). Besides, from the perspective of molecular function, a previous study has reported that the overexpression of FLI1 accelerates embryonic stem cell differentiation into endothelial cells[59]. It is also reported that EGR1 is widely expressed in many cell types and participates in important physiological processes, such as cell proliferation, differentiation, invasion, and apoptosis[60]. These collective findings indicate a high likelihood of the transcription factor EGR1 regulating FLI1 expression by binding to its promoter region. This also implies that SCALD possesses the capacity to predict previously undetected regulatory relationships. Given that the regulatory relationships between molecules dynamically change across different states, our approach aids in further refining condition-specific regulatory relationships using state-specific data.

## SCALD contributes to identify tumor process-associated transcription factors.

The transformation from chronic inflammation to cancer has long been a topic of significant interest, as it is crucial for understanding the mechanisms of cancer development and beneficial for taking early intervention. Existing studies have demonstrated that the regulatory mechanisms indeed change and promote the state transition of disease progression[61]–[63]. In this study, taking colon disease as an example, we apply SCALD to identify the main regulatory distinctions among normal tissue, inflammatory bowel diseases (IBD), and colorectal carcinomas (CRC). To do this, we first collect raw gene expression data GSE4183 from GEO database[64]–[66], which comprises 8 normal samples, 15 inflammatory bowel diseases (IBD) samples and 15 colorectal carcinomas (CRC) samples, respectively.

Through differential expression analysis between the normal and CRC tissues using the Wilcoxon test, and with a significance threshold set at p-value < 0.001, we identify 669 differentially expressed genes, comprising 34 TFs and 635 TGs, for further investigation. Then, we use SCALD to infer a gene regulatory network for each group. We assume that the regulatory relationships exhibiting monotonic increase or decrease in strength have a crucial function in colon cancer progression. By examining relations for significant monotonic changes with a difference greater than 0.15 between adjacent stages, we identified 73 cancer progression-related regulatory relationships, which constitute a cancer-related regulatory network, as shown in Fig. 6. Among all the significantly altered regulatory relationships, we observe that some of them share the same transcription factors, leading us to speculate that transcription factors having high degree in cancer-related regulatory network, such as *ETV3, PRRX1, ZSCAN22, MIER3 and SOX7*, play key roles in the cancer development. Moreover, these predicted cancer-related transcription factors we identified are consistent with existing studies. For example, *ETV3*, a member of the E-26 (ETS) family of transcription factors, has been shown to be a potential prognostic marker for colorectal cancer[67]. *PRRX1* has been revealed as a significant catalyst in epithelial-mesenchymal transition, contributing to colorectal cancer metastasis and unfavorable prognosis[68]. Besides, Peng et al. provided evidence that the up-regulation of *MIER3* significantly suppresses CRC cell proliferation, migration, and invasion[69]. Zhang et al. discovered that *SOX7* is usually down-regulated in human colorectal cancer cell lines and primary colorectal tumor tissues[70].

In conclusion, the gene regulatory relationships we identified that highly associated with CRC development offer valuable insights into unraveling the origins, progression, and transformative processes of tumors. This holds the promise of establishing a new theoretical foundation for future cancer treatment and prevention strategies.

## Discussion

In contrast to the majority of existing studies that infer biological networks based on generalized correlation, our approach emphasizes the importance of causality. While several methods for causal structure inference do exist, they are predominantly under the Directed Acyclic Graph (DAG) assumption. This assumption, however, limits their direct applicability to biological networks due to the presence of feedback loops in real biological systems. To address this, we propose SCALD, a causal graph structure learning framework specifically designed for scenarios with feedback loops. SCALD employs a combination of deep learning and structural equation models to initially discern the associations among variables. We then delve into the topological properties of loops, through introducing symbolic information that signifies promotion or inhibition relationships and considering the conditions of stable systems on cross-sectional data, to infer the causal network structure. This enables SCALD to infer a wide range of biological networks, including but not limited to regulatory and signaling networks.

SCALD has proven effective in inferring both gene regulatory networks and signaling networks when applied to transcriptomics and proteomics data, respectively. Beyond its alignment with ground truth, we further validate the reliability of SCALD in inferring regulatory relationships by observing the impact of perturbed Transcription Factors (TFs) on target Transcriptional Genes (TGs) and the binding degree of TFs in the TG's promoter region. Through ChIP-seq data validation, we demonstrate that the regulatory relationships identified by SCALD, which are not present in the ground truth, offer a valuable augmentation to the existing, incomplete set of regulatory relations.

The accurate inference of biological networks enable us to understand the intricate interactions among various components in the complex biological systems. This understanding helps the identification of key factors of system through topological analysis, discovery of potential drug targets by estimating and analyzing the perturbation effects of certain genes, and so on. In this study, we employ SCALD to deduce biological networks at different stages of transition from chronic bowel inflammation to intestinal cancer. By analyzing the topological alterations in each network throughout the transformation process, we aim to predict potential driver genes and pivotal pathways involved in the evolution of intestinal cancer. Furthermore, we anticipate that SCALD will facilitate the identification of key drivers and regulatory mechanisms in the progression of other complex diseases.

The success of SCALD is attributed to the construction of biological networks from a causal relationship perspective and effective characterization of the properties of widely existing feedback loops in biological networks. The result analysis also proves that SCALD has apparent advantages in predicting the feedback loops within biological network. Simultaneously, we also face common challenges when constructing networks. This approach is limited by the relatively high computational complexity of loop constraint. Therefore, in future work, we plan to employ superior nonlinear dimensionality reduction techniques that preserve the neighborhood structure of high-dimensional space in a lower-dimensional representation, such as Isometric Mapping, Locally Linear Embedding, Factor Graph Analysis, and so on. Particularly when dealing with a small number of samples, we contemplate the use of generative models to augment the sample size, thereby meeting the requirement of a large sample volume for training neural networks.

## Method

### Data collection and processing.

To evaluate the performance of SCALD in GRN inference, we collect benchmark datasets *E.coli* and *S.cerevisiae* (*S.cere*) from the DREAM5 challenge, along with scRNA-seq data of human embryonic stem cells (hESC) from BEELINE benchmark repository [36]. The ground truth labels for regulation relationships in the E.coli dataset are derived from experimentally validated interactions sourced from RegulonDB[71]. Meanwhile, the ground truth for the S.cerevisiae dataset is obtained from a high-confidence set of interactions, which are supported by genome-wide transcription factor binding data [72] and evolutionarily conserved binding motifs [73]. To comprehensively assess the GRN inference performance of SCALD on single cell data, we have conduct six experiments by using two subsets of data (containing either the 500 or 1,000 most variable genes) under three different types of ground-truth networks (Cell-type-specific, Nonspecific, and STRING). STRING data are sourced from functional connectivity within the STRING database[37], while Cell-type-specific and Nonspecific data are derived from BEELINE benchmark repository[36]. In detail, Cell-type-specific are generated using ChIP-seq data obtained from ENCODE, ChIP-Atlas and ESCAPE databases. The NonSpecific are generated using nonspecific ChIP-seq data according to DoRothEA[74], RegNetwork[75], and TRRUST[76] databases.

To evaluate the performance of SCALD in signaling network inference, we have collect the Sachs dataset comprises the simultaneous measurements of 11 phosphorylated proteins and phospholipids derived from 7466 primary immune system cells for both general and specific molecular interventions [28], [77] . Based on the probability values, we have selected the top 20 predicted edges to evaluate the performance of SCALD.

To further validate the trustworthiness of regulatory relationships predicted by SCALD, we have collected perturbation data. According to our inference results, which indicate ELK1 and NFATC3 are two important transcription factors for CD4+ T cell, we have obtain CD4+ T cell expression data from the GEO database(GSE46333) with and without ELK1 and NFATC3 perturbation. Through comparing the differential expression level of their downstream target genes between the perturbed and unperturbed groups, we can determine whether the regulatory relationships predicted by SCALD is trustworthy. We use p-value< 0.05 and absolute log2 fold-change> 1 as the cutoff in the differential expression analysis.

Regarding regulatory relationships predicted by SCALD but not present in the established ground truth, we extend our investigation by conducting analyses of ChIP-seq data. The ChIP-seq data of hESC are downloaded from ENCODE database.

To investigate changes in regulatory networks throughout tumor development, we have obtained mRNA expression data for normal, Inflammatory Bowel Disease (IBD), and Colorectal Cancer (CRC) states from the public microarray dataset GSE4183. When multiple probes correspond to one gene, we select the probe with the highest mean value and use its expression value to represent the gene. We retrieve the list of transcription factors (TFs) from AnimalTFDB 3.0[78], and exclude tissue-specific genes of other tissues, except for the colon, from TRRUST. Following a differential expression analysis of normal and CRC samples using a t-test with a p-value <0.001, we identified 536 differentially expressed genes, including 30 transcription factors and 506 target genes, for further analysis. All the above data are publicly available.

## Nonlinear Structural Causal Model.

We first construct a basic causal regulatory network skeleton based on Structure Equation Model. Given $\mathbf{X} \in \mathbb{R}^{n \times d}$ containing $n$ i.i.d. observed data samples $x = (x_1, x_2, \ldots, x_d)$, our goal is to determine an adjacency matrix $\mathbf{A} \in \mathbb{R}^{d \times d}$ representing a directed causal graph structure, where if $A_{ij} \neq 0$, then there exists a directed edge $i \to j$, representing variable (gene) $i$ regulates variable (gene) $j$. Taking a linear regulatory relationship as an example, it is mathematically represented as $\mathbf{X} = \mathbf{XA} + \mathbf{E}$, where $\mathbf{E} \in \mathbb{R}^{d \times d}$ denotes the noise matrix. The estimation of the $j$-th feature variable $x_j$ in vector $x$ can be obtained through the equation: $x_j = \sum_{i=0, i \neq j}^{d} x_i \mathbf{A}_{ij} + e_j$, where $e_j$ represents the noise variable for all parent nodes of $x_j$. We assume that the noise term $e_j$ related to node $j$ is independent of $j$'s parent nodes, i.e., $e_j \perp pa(x_j)$. Additionally, to meet with the constraint of preventing self-cycles in the circular SEM [31], we assign the diagonal elements of **A** as $\mathbf{A}_{ii} = 0$.

Considering the regulation are usually nonlinear, we introduce a fully connected feedforward neural network $g(.)$ to model the relationships also under the assumption of noise independence.

The nonlinear model is described as $\mathbf{X} = g(\mathbf{XA}) + \mathbf{E}$, where $\mathbf{A}_{ij}$ represents the probability of the existence of the directed edge $x_i \rightarrow x_j$. Besides, we apply a constraint on the adjacency matrix **A** to guarantee stable feedback regulation, thereby eliminating cascading amplification or attenuation loops. As a result, we use Ordinary Least Squares (OLS) to minimize the difference between observed and predicted values, and the loss function is expressed as:

$$f(\mathbf{A}) = (\mathbf{X} - g(\mathbf{XA}))^2 \tag{1}$$

Thus, we seek to solve:

$$\min_{\mathbf{A}} f(\mathbf{A}) + \rho_l \|\mathbf{A}\|$$
$$s.t. \; \mathbf{A} \text{ satisfies the stability of loops} \Leftrightarrow h(\mathbf{A}) = 0 \tag{2}$$

We include an L1 regularization term and a constraint on **A**, $\rho_l \|\mathbf{A}\|$ to enhance sparsity of the graph, and also add a constraint function $h(\mathbf{A}) = 0$ to ensure that the causal graph satisfies the systematic stability of loops. The property of stable loops and construction for $h(\mathbf{A}) = 0$ are presented in Section 4.3.

### Constraint for Stable Loops.

Here, we introduce sign attributes for each edge, $\mathbf{A}_{ij}$, to assist in assessing the loops stability in different situations, and propose corresponding constraint formulas that can be optimized continuously. We define a matrix $\mathbf{S} \in \mathbb{R}^{d \times d}$ to represent the sign relationships, which utilizes the Spearman correlation coefficient to quantify the similarity between nodes and determine the signs of their association, positive or negative. Specifically, $\mathbf{S}_{ij} \in [-1, 1]$, $\mathbf{S}_{ij} > 0$ or $\mathbf{S}_{ij} < 0$ denote a positive or negative correlation between nodes *i* and *j*, respectively, while $\mathbf{S}_{ij} = 0$ represent there is no significant association between nodes *i* and *j* under hypothesis test (p-value < $\varepsilon$ ). Here, the cutoff depends on the number of samples.

$$\varepsilon = \begin{cases} 0.05 & , \text{if } n \leq 20 \\ 0.001 & , \text{if } 20 < n \leq 100 \\ 0.0005 & , \text{if } n > 100 \end{cases} \tag{3}$$

On the basis of the sign matrix **S** and the adjacency matrix **A** (as described in Section 4.2), we generate two new mono-signed adjacency matrixes $\mathbf{A}_{pos}, \mathbf{A}_{neg} \in \mathbb{R}^{d \times d}$, which are used for constraining the learning to ensure the stable biological feedback.

$$\begin{aligned}
(\mathbf{A}_{pos})_{ij} &= \begin{cases} \mathbf{A}_{ij}, & \text{if } \mathbf{S}_{ij} > 0 \text{ and } p_{ij} > \varepsilon \\ 0, & \text{if } \mathbf{S}_{ij} \leq 0 \text{ or } p_{ij} \leq \varepsilon \end{cases} \\
(\mathbf{A}_{neg})_{ij} &= \begin{cases} 0, & \text{if } \mathbf{S}_{ij} > 0 \text{ and } p_{ij} > \varepsilon \\ \mathbf{A}_{ij}, & \text{if } \mathbf{S}_{ij} \leq 0 \text{ or } p_{ij} \leq \varepsilon \end{cases}
\end{aligned} \quad (4)$$

Further, the constraint $h(\mathbf{A}) = 0$ presented in Formula (2) can be regarded as a combination of the following two constraint terms,

$$\begin{aligned}
h_1(\mathbf{A}) &= tr((\mathbf{I} + \mathbf{A}_{pos})^d) - d = 0 \\
h_2(\mathbf{A}) &= tr((\mathbf{I} + \mathbf{Q})^{d/2}) - d = 0
\end{aligned} \quad (5)$$

where h1(A) is employed to restrict loops composed solely of positive edge, and h2(A) limits loops consisting of an even number of negative edges along any with number of positive edges.

The proof is as follows:

(i) For an unsigned directed adjacency matrix **A**, if $\mathbf{A}_{ij}^2 = \sum_{k=1}^{d} \mathbf{A}_{ik} \mathbf{A}_{kj} \neq 0$, it indicates that node *i* can indirectly affect node *j* through an intermediate node. By analogy, when $\mathbf{A}_{ij}^{k+1} \neq 0$, node *i* can reach node *j* through a path of length *k*. Therefore,

$$\begin{aligned}
\text{node } i \text{ is not in a cycle} &\Leftrightarrow \sum_{k=1}^{d} \mathbf{A}_{ii}^k = \sum_{k=1}^{d} tr(\mathbf{A}^k) = 0 \\
&\Leftrightarrow tr(\mathbf{I} + c_1 \mathbf{A} + c_2 \mathbf{A}^2 + c_3 \mathbf{A}^3 + \ldots + c_d \mathbf{A}^d) - d = tr((\mathbf{I} + \mathbf{A})^d) - d = 0,
\end{aligned} \quad (6)$$

where $c_i$ is an arbitrary non-zero coefficient.

(ii) For a signed directed adjacency matrix, if the variable B and A exhibit the same trend of change, the link from variable A to B is marked with a "+". Otherwise, if the variable B changes in the opposite direction to A, the link from A to B should be labeled as "-". We refer to this type of graph as a Causal Loop Diagram (CLD), where the loop can be divided into two main categories with following characteristics, as depicted in Fig. 7:

**Reinforcing Loops**: As shown in Fig. 7A(a-d), if there is no or even negative edges in the loop, the signal transmitted back to the original node shows a cascade of amplification or reduction at a particular time point or state.

**Stable Loops:** When the number of negative edges in the loop is odd, the signal transmitted back to the original node stabilizes the system through feedback regulation, as shown in Fig. 7A(e-h).

To maintain the stability of the whole biological system, it is posited that stable loops ought to constitute the system's primary structure. To achieve this, we eliminate the reinforcing loops in the system as much as possible. Hence, considering both the number of negative edges and the physical meaning of multiplying adjacency matrices (Formula (6)), we analyze topological properties of these two types of loops and design corresponding constraints to limit reinforcing loops. When there are no negative edges in the loop, we extract its positive edges and subject them to a directed acyclic constraint:

$$\mathbf{I} + c_1\mathbf{A}_{pos} + c_2\mathbf{A}_{pos}^2 + c_3\mathbf{A}_{pos}^3 + ... + c_d\mathbf{A}_{pos}^d = (\mathbf{I} + \mathbf{A}_{pos})^d$$
$$tr((\mathbf{I} + \mathbf{A}_{pos})^d) - d = 0 \tag{7}$$

where $c_i$ is an arbitrary non-zero coefficient, similar to the formulation in Formula (6). On the other hand, when there is an even number of negative edges within a loop, the edges in the loop can be considered as having any number (including 0) of positive edges inserted along any negative edge paths.

In terms of how to represent such a loop, we start with all positive edges in it:

$$\mathbf{B} = \mathbf{I} + c_1\mathbf{A}_{pos} + c_2\mathbf{A}_{pos}^2 + c_3\mathbf{A}_{pos}^3 + ... + c_d\mathbf{A}_{pos}^d = (\mathbf{I} + \mathbf{A}_{pos})^d \tag{8}$$

where $\mathbf{B} \in \mathbb{R}^{d \times d}$ represents the connection between node i and node j only through positive edges. That is, $\mathbf{B}_{ij} \neq 0$ indicates that i can connect to j only through positive edges. On this basis, we consider how to identify the loops that consists of an arbitrary number of positive edges and an even number of negative edges. In the simplest case, for example, we consider there are only two negative edges in the loop, which is denoted as two negative edges connecting with an arbitrary number of positive edges, i.e, $\mathbf{Q} = \mathbf{A}_{neg}\mathbf{B}\mathbf{A}_{neg}\mathbf{B}$. In this situation, the constraint is $tr(\mathbf{Q}) = 0$. Furthermore, when there are four negative edges in the loop, it is denoted as $\mathbf{A}_{neg}\mathbf{B}\mathbf{A}_{neg}\mathbf{B}\mathbf{A}_{neg}\mathbf{B}\mathbf{A}_{neg}\mathbf{B}$ and the corresponding constraint is $tr(\mathbf{Q}^2) = 0$. Following this pattern, we can easily extend it to cases where there are arbitrary even number of negative edges:

$$tr(\mathbf{Q} + \mathbf{Q}^2 + \mathbf{Q}^3 + ... + \mathbf{Q}^{d/2}) = 0 \tag{9}$$

Similar to Formula (6) the constraint of this condition can be expressed as:

$$tr(\mathbf{I} + \mathbf{Q} + \mathbf{Q}^2 + \mathbf{Q}^3 + ... + \mathbf{Q}^{d/2}) - d = 0 \Leftrightarrow tr((\mathbf{I} + \mathbf{Q})^{d/2}) - d = 0 \tag{10}$$

To provide a clear illustration, we present an example of a loop with an even number of negative edges, as depicted in Fig. 7B, to show that our proposed constraint can be applied to any loops composed of an arbitrary even number of negative edges.

Therefore, the constraint for stable loops consists of two parts: $tr((\mathbf{I}+\mathbf{A}_{pos})^d) - d = 0$ and $tr((\mathbf{I}+\mathbf{Q})^{d/2}) - d = 0$, and the constraint formula $h(\mathbf{A}) = 0$ presented in Formula (2) can be presented as a combination of the following two constraint terms:

$$h_1(\mathbf{A}) = tr((\mathbf{I}+\mathbf{A}_{pos})^d) - d = 0$$
$$h_2(\mathbf{A}) = tr((\mathbf{I}+\mathbf{Q})^{d/2}) - d = 0 \qquad (11)$$

□

## Optimization of SCALD.

Regarding the final loss function as follows:

$$\min_{\mathbf{A}} f(\mathbf{A}) + \rho_l \|\mathbf{A}\|^2$$
$$s.t.\ h_1(\mathbf{A}) = 0, h_2(\mathbf{A}) = 0 \qquad (12)$$

we use the augmented Lagrangian method to solve the constrained optimization problem by transforming it into an equivalent unconstrained problem. First, we transform the loss function to its dual form:

$$D(\lambda_1, \lambda_2) = \min_{\mathbf{A}} L(\mathbf{A}, \lambda_1, \lambda_2; \rho_1, \rho_2, \rho_l)$$
$$L(\mathbf{A}, \lambda_1, \lambda_2; \rho_1, \rho_2, \rho_l) = f(\mathbf{A}) + \rho_l \|\mathbf{A}\|^2 + \lambda_1 h_1(\mathbf{A}) + \lambda_2 h_2(\mathbf{A}) + \frac{\rho_1}{2}(h_1(\mathbf{A}))^2 + \frac{\rho_2}{2}(h_2(\mathbf{A}))^2 \qquad (13)$$

where $L(\mathbf{A}, \lambda_1, \lambda_2; \rho_1, \rho_2, \rho_l)$ is the augmented Lagrangian function. Then we solve the dual problem as follows:

$$\max_{\lambda_1, \lambda_2} D(\lambda_1, \lambda_2) \qquad (14)$$

The derivatives of parameters $\lambda_1, \lambda_2$ is obtained by $\nabla(\lambda_1) = h_1(\mathbf{A}^*), \nabla(\lambda_2) = h_2(\mathbf{A}^*)$, where $\mathbf{A}^*$ is the local minimizer of the $D(\lambda_1, \lambda_2)$ at $\lambda_1, \lambda_2$, i.e., $D(\lambda_1, \lambda_2) = L(\mathbf{A}^*, \lambda_1, \lambda_2; \rho_1, \rho_2, \rho_l)$. Thus, we update $\lambda_1, \lambda_2$ using:

$$\lambda_1 \leftarrow \lambda_1 + \rho_1 h_1(\mathbf{A}^*)$$
$$\lambda_2 \leftarrow \lambda_2 + \rho_2 h_2(\mathbf{A}^*) \qquad (15)$$

## Configuration of Edge Signs.

In the proposed is a nonlinear model, $\mathbf{A}_{ij}$ represents the probability that the directed edge $x_i \rightarrow x_j$ exists, rather than the weight of the edge. Therefore, we constrain $\mathbf{A}_{ij} \in (0,1)$ using sigmoid(**A**). Additionally, our method allows for loops in the graph, but self-loops are not allowed. Hence, the diagonal elements of A are set to 0. In GRNs, transcription factors (TFs) regulate the expression of target genes (TGs)，but the reverse is not true. Thus, we collect TF information as prior knowledge and remove all connections of the form TG→TF for building original association during the prediction process. Additionally, as small value of $|\mathbf{S}_{ij}|$ and large p-value indicate weak relationships between nodes *i* and *j*, we use the p-value $p_{ij}$ which is corresponding to Spearman correlation to select the highly reliable pats of **A**, i.e. $\mathbf{A}_{pos}, \mathbf{A}_{neg}$,

$$\begin{cases} (\mathbf{A}_{pos})_{ij} = \mathbf{A}_{ij}, (\mathbf{A}_{neg})_{ij} = 0, & if \ \mathbf{S}_{ij} > 0 \text{ and } p_{ij} \leq \varepsilon \\ (\mathbf{A}_{neg})_{ij} = 0, (\mathbf{A}_{neg})_{ij} = \mathbf{A}_{ij}, & if \ \mathbf{S}_{ij} \leq 0 \text{ and } p_{ij} \leq \varepsilon \end{cases} \quad (16)$$

## Evaluation metrics.

For the adjacency matrix A with values denoting edge existence probability, we sort the directed edges in descending order. Three common metrics, Early Precision Ratio (EPR), Area Under the Precision-Recall Curve (AUPR), and Area Under the Receiver Operating Characteristic Curve (AUROC), are employed to evaluate the inference performance.

EPR quantifies the accuracy of the top-k predicted edges by comparing them to a baseline established through randomly selecting *k* edges, repeated across multiple iterations. Here, *k* corresponds to the actual number of edges present in the ground truth data. To reduce any biases introduced by random selection of *k* edges, we performed this process 100 times and utilized the mean as the final result.

AUPR and AUROC are metrics used to evaluate the performance of binary classification models. They calculate the area under the Precision-Recall curve and the ROC curve, respectively. Precision-Recall curve is plotted based on different thresholds of probabilities predicted by the model, illustrating the Precision and Recall at varying thresholds. These curves offer insights into model performance at different levels of sensitivity and specificity. The formulas for TPR, FPR, Precision, and Recall are as follows:

$$Recall = TPR = \frac{TP}{TP + FN} \quad (17)$$

$$Precision = \frac{TP}{TP + FP} \quad (18)$$

$$FPR = \frac{FP}{TN + FP} \quad (19)$$

where True Positive (TP) represents the number of edges that both exist and are predicted to exist; False Positive (FP) refers to the number of edges that do not exist but are predicted to; True Negative (TN) denotes the number of edges that neither exist nor are predicted to.

## Acknowledgments

The authors thank funding support from the National Natural Science Foundation of China (No. 62372210, U2341229), Natural Science Foundation of Jilin Province(20240101025JJ), and Jilin University organized scientific research project (No. 45123031J004).

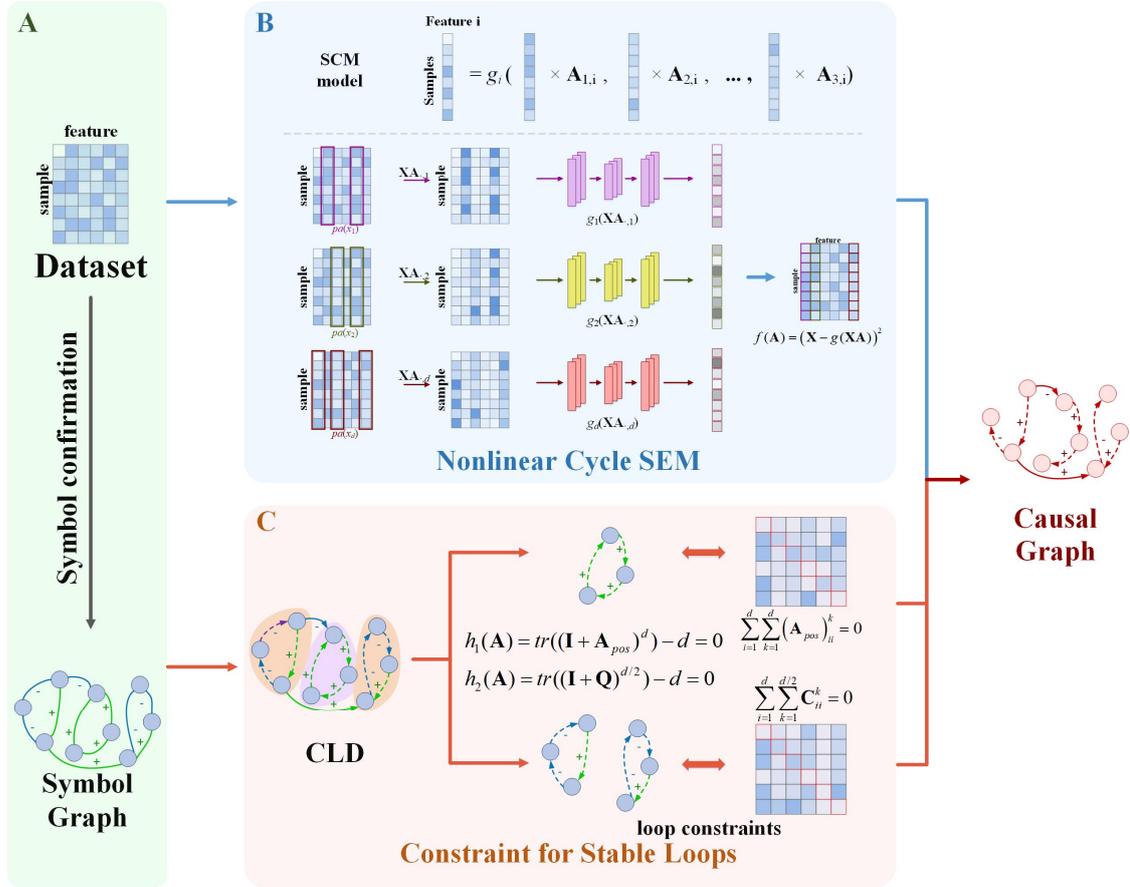

**Fig. 1 Overview of SCALD.** (A) Given $\mathbf{X} \in \mathbb{R}^{n \times d}$ containing *n* i.i.d. observed data samples, each one comprising d genes, we first determine the signs of gene-gene association based on Spearman correlation, and then obtain a symbolic graph according to correlation significance. (B) Learning an unsigned rough directed graph using the non-linear structure equation model. (C) We design two constraints trying to avoiding unstable feedback in cross-section data when the number of negative edges in the loop is zero and even, respectively.

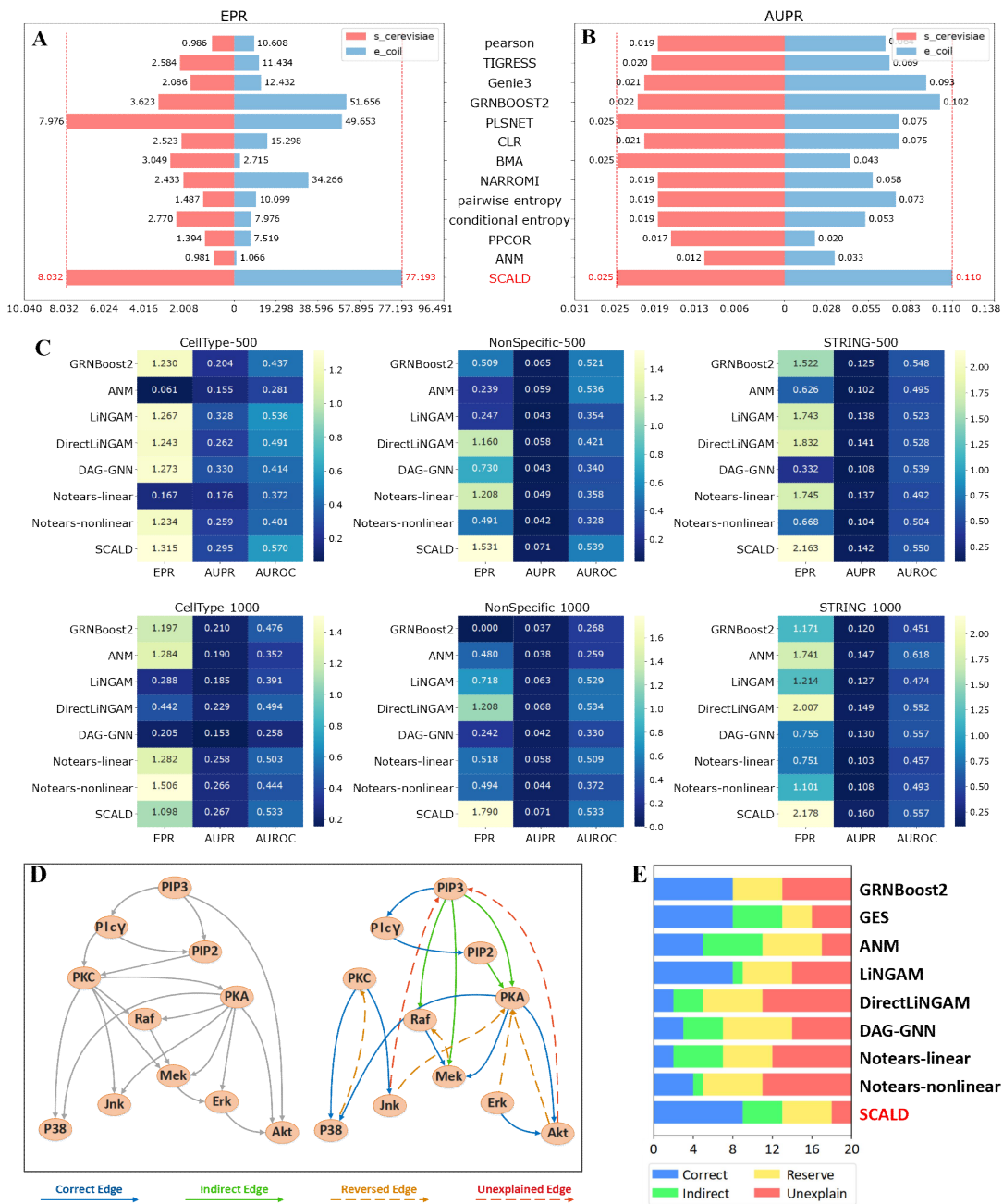

**Fig. 2 Summary of the GRN and signaling network prediction performance.** (A) The performance of GRN inference on *E.coli* (blue) and *S.cerevisiae* (*S.cere*) (red) datasets in term of EPR. The red dashed line represents the value of SCALD. (B) The performance of GRN inference on *E.coli* (blue) and *S.cerevisiae* (*S.cere*) (red) datasets in term of AUPR. The red dashed line represents the value of SCALD. (C) The results of six experiments (CellType-500, CellType-1000, NonSpecific-500, NonSpecific-1000, STRING-500, and STRING-1000) on hESC scRNA-seq data are presented under three evaluation metrics: EPR, AUPR, and AUROC. EPR measures the accuracy of the top-k predicted edges by contrasting them with a baseline established through random selection of k edges, repeated across

multiple iterations. AUPR and AUROC compute the area under the Precision-Recall curve and the ROC curve, respectively. For each dataset, the color scales between 0 and 1 by the min-max scale based on the EPR value.

(D) The ground truth network structure of the PI3K-AKT signaling pathway and the predicted network generated by SCALD through the Sachs perturbation dataset are depicted. (E) The statistical results of nine causal structure learning methods on estimating the directed regulatory relationship of the Sachs dataset are presented. Each method selected the top 20 edges with the highest probabilities. The solid blue lines represent the correctly predicted edges, the solid green lines represent indirectly connected edges, the direction of yellow dashed lines are contrary to the ground truth, and the red dashed lines are either not present in the ground truth or lack clear evidence to substantiate their existence

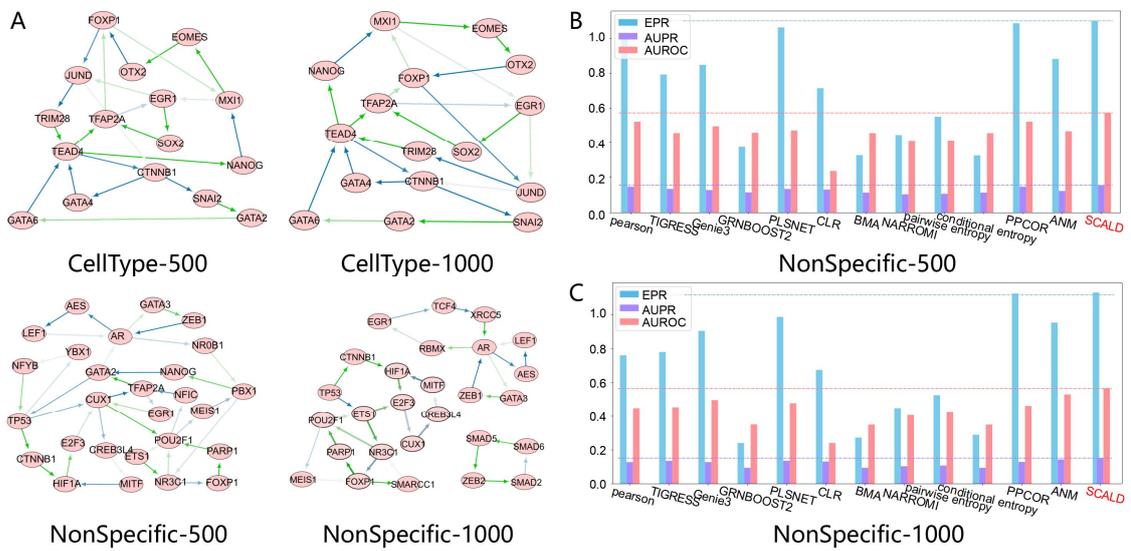

**Fig. 3 Statistics of stable loops and the performance of the feedback loops inference.** (A) Illustration of stable loops in four ground-truth networks(CellType-500, CellType-1000, NonSpecific-500, and NonSpecific-1000) of scRNA-seq data. Green edges denote positive regulations and blue edges denote negative regulations. (B) The performance of 13 methods on the NonSpecific-500 dataset is evaluated under EPR, AUPR, and AUROC metrics. Blue represents the EPR metric, purple signifies the AUPR metric, and red denotes the AUROC metric. The three dashed lines (blue, purple, and red) correspond to the results (EPR, AUPR, and AUROC) of SCALD, facilitating easy comparison with other algorithms. (C) The performance of 13 methods (Pearson, TIGRESS, Genie3, GENBOOST2, PLSNET, CLR, BMA, NARROMI, pairwise entropy, conditional entropy, PPCOR, ANM, and SCALD) on the NonSpecific-1000 dataset is evaluated under EPR, AUPR, and AUROC metrics.

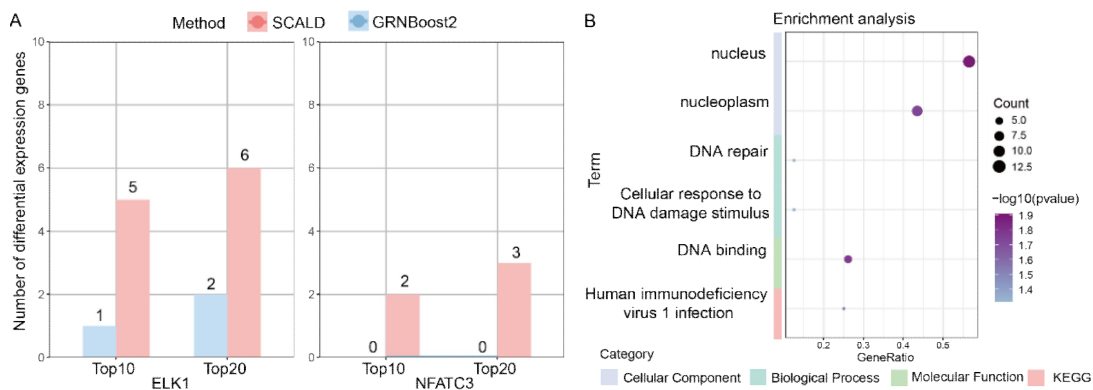

**Fig. 4 Validation of GRN with transcription factor perturbation data.** (A) Statistics for the number of differentially expressed genes of pre- and post-perturbation, contained in the top 10 and 20 target genes of NFATC3 and ELK1 predicted by SCALD. (B) GO and KEGG functional enrichment analysis on the top30 target genes of NFATC3.

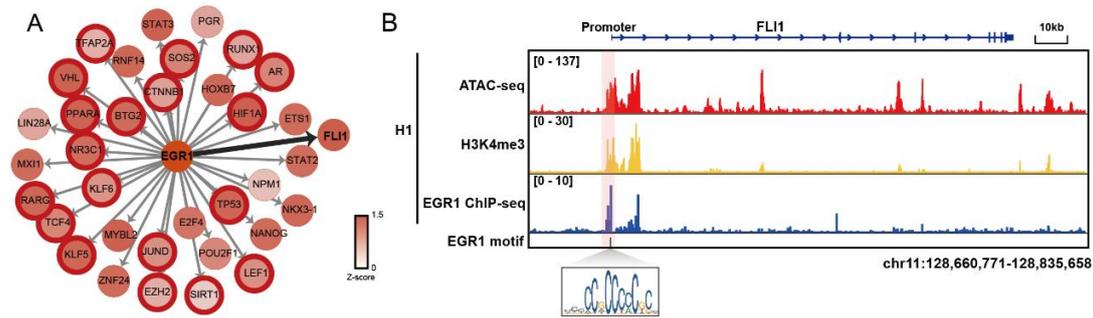

**Fig. 5 Validation of EGR1-FLI1 regulatory relationship with epiginomic data.** (A) In the subnetwork, node colors signify the strength of the predicted relationship between the node and EGR1, represented by normalized Z-scores. Darker colors correspond to higher scores. Nodes with bolded borders indicate regulations that exist in the ground truth. In this context, we specifically focus on the relationship between EGR1 and FLI1, which is not represented in the ground truth. (B) Integrative Genomics Viewer snapshots with experimental data of EGR1-FLI1. The signal distribution of TF EGR1 ChIP-seq data, H3K4me3 ChIP-seq data, and ATAC-seq data in the *FLI1* promoter region of the H1 cell line.

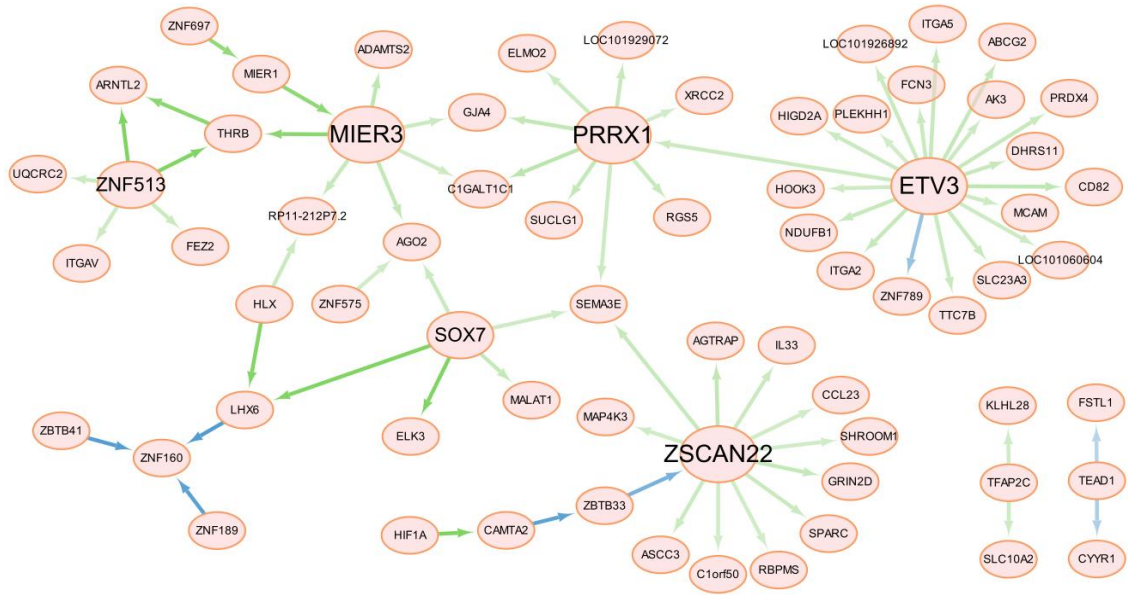

**Fig. 6 The inferred core Gene Regulatory Networks that drive the progression from chronic colon inflammation to colon cancer.** Regulatory relationships that are potentially associated with promoting the transformation of chronic inflammation into cancer are depicted. Each edge in this figure represents regulatory relationships that change as chronic inflammation transitions into cancer, exhibiting a monotonic trend during this transformation. Green edges represent interactions that strengthen with cancer progression, while blue edges indicate a weakening relationship. The size of a node represents its level of involvement in differential regulation. Transcription factors represented by larger nodes indicate a greater degree of connectivity in the network, presumably suggesting a closer association with driving the transformation from inflammation to cancer.

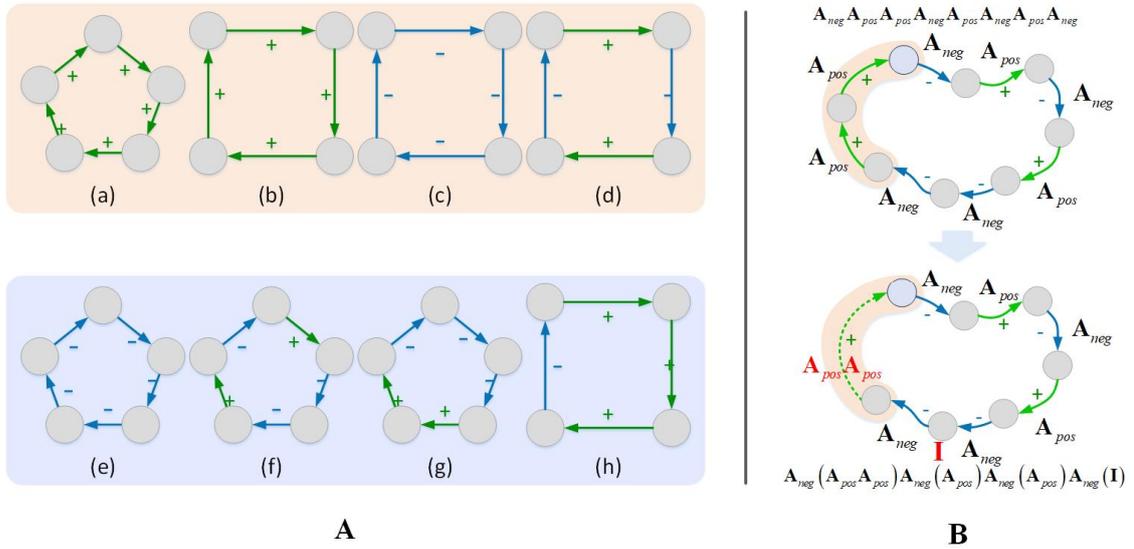

**Fig. 7 Loop classification and constrained examples**. (A) We provide several examples of reinforcing loops (a-d) and stable loops (e-h). (B) The upper part of the figure represents a loop in the directed graph, while the lower part displays the corresponding shape obtained by connecting all adjacent positive edges, where the item in brackets can be found in the lower part, such as $(\mathbf{A}_{pos}\mathbf{A}_{pos})$, $(\mathbf{A}_{pos})$, and $(\mathbf{I})$. It can be observed that the segment within the parentheses can be expressed by one of the terms in B, irrespective of the number of adjacent positive edges in the loop

# Supplementary

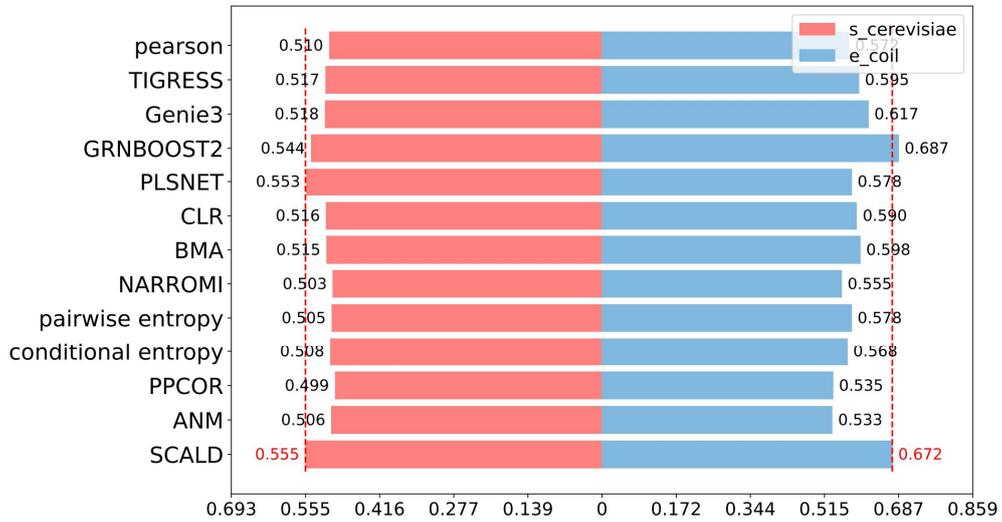

**Fig. S1 The regulatory performance of SCALD and other 12 methods on Dream5 dataset under AUROC.** The baselines include pearson, TIGRESS, Genie3, GENBOOST2, PLSNET, CLR, BMA, NARROMI, pairwise entropy, conditional entropy, PPCOR, and ANM. The red bars are the performances of GRN inference on *S.cere*, and the blue bars are the performances of GRN inference on *E.coli*. The red dashed lines correspond to the results of SCALD for easy comparison with other algorithms.

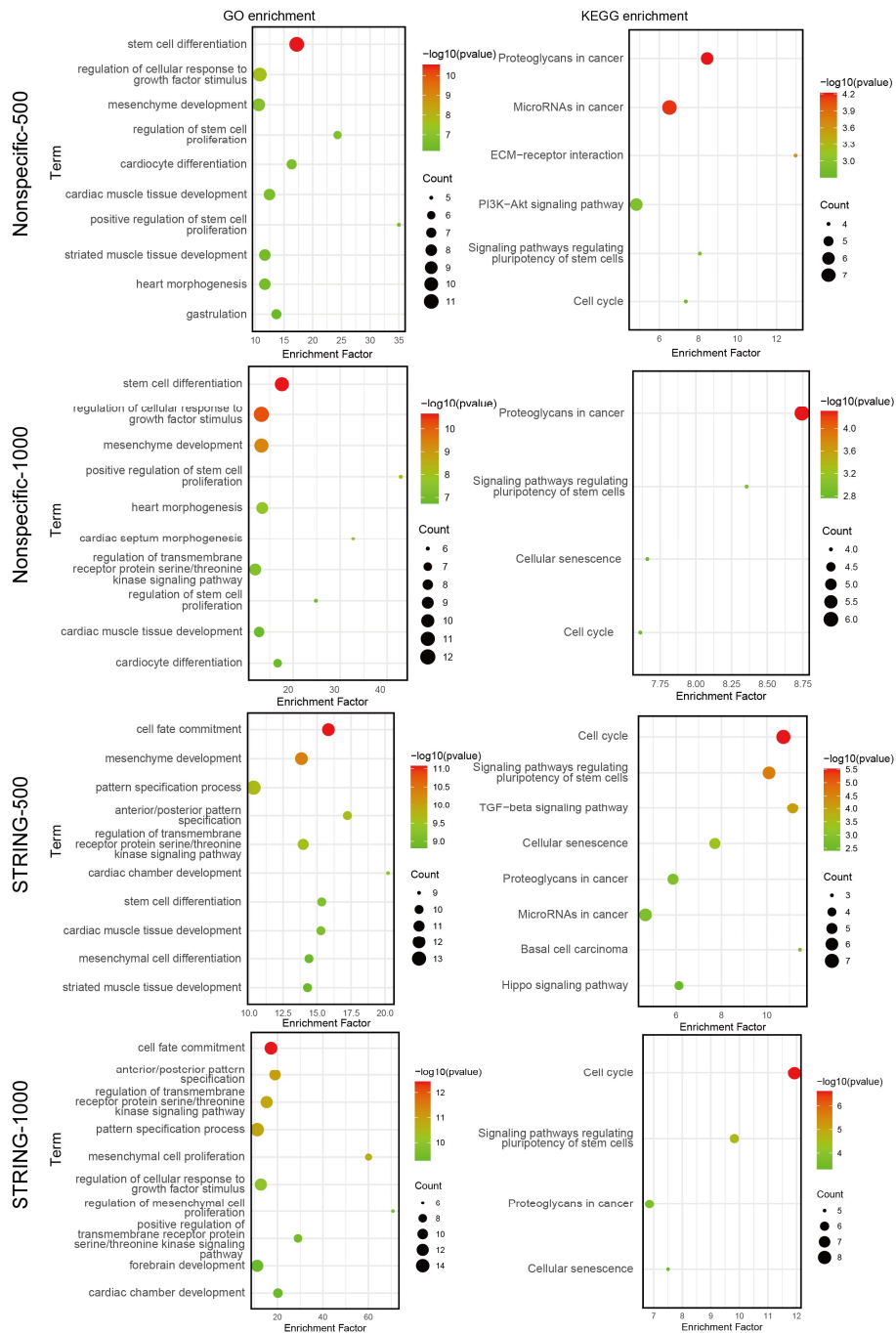

**Fig. S2 Bubble plots of KEGG and GO enrichment analyses are presented for nodes predicted to be in the top 50 based on centrality from the predicted GRN derived from NonSpecific-500, NonSpecific-1000, STRING-500, and STRING-1000 datasets**. The y-axis represents either GO terms or KEGG terms. The x-axis signifies the Enrichment Factor score, which indicates the ratio of genes annotated to the pathway in the input gene set relative to the proportion of human genes annotated to that same pathway. A higher Enrichment Factor score denotes more significant enrichment. The size of the bubbles corresponds to the count of genes annotated to that pathway in the input gene set

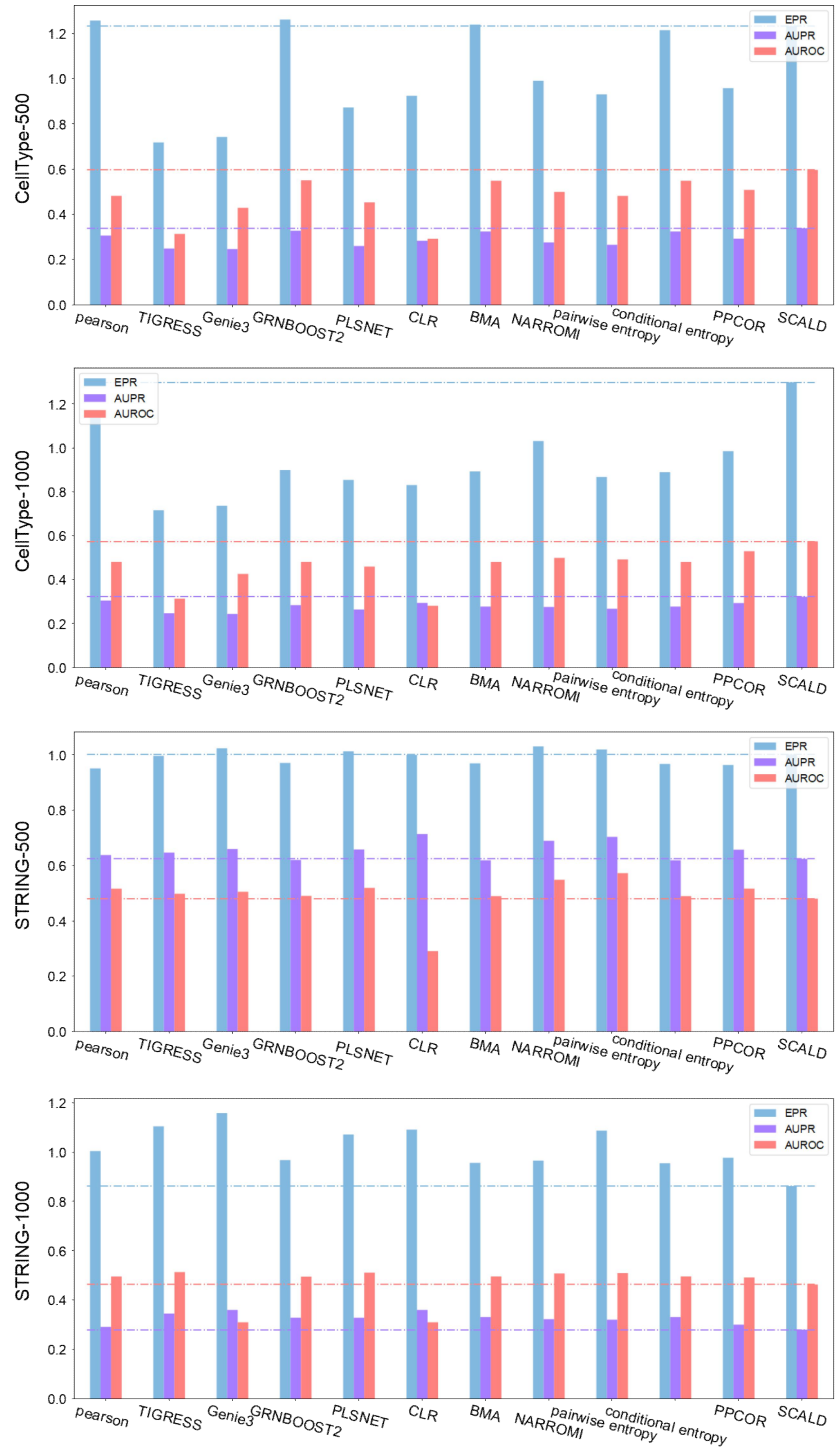

**Fig. S3 Performance of different methods (pearson, TIGRESS, Genie3, GENBOOST2, PLSNET, CLR, BMA, NARROMI, pairwise entropy, conditional entropy, PPCOR, and SCALD) on CellType-500, CellType-1000, STRING-500 and STRING-1000 datasets under EPR, AUPR and AUROC metrics.** The red bars represent the performance of GRN inference on S.cerevisiae, while the blue

bars denote the performance of GRN inference on E.coli. Within these, the blue bar signifies the EPR metric, the purple bar corresponds to the AUPR metric, and the red bar represents the AUROC metric. The three dashed lines, in blue, purple, and red, correspond to the EPR, AUPR, and AUROC results of SCALD, respectively, facilitating easy comparison with other algorithms.

Table S1 The count of enrichment pathways observed in six scRNA-seq experiments of hESC, along with the related researches on these pathways in hESC.

| | Enrichment Pathways | Count | Link |
|---|---|---|---|
| GO | Stem cell differentiation | 5 | Pan et al.[38] |
| | pattern specification process | 3 | Muncie et al.[39] |
| | cell fate commitment | 3 | Tatapudy et al.[40] |
| | mesenchyme development | 3 | Vasanthan et al.[41] |
| KEGG | Signaling pathways regulating pluripotency of stem cells | 6 | Okita et al.[46] |
| | Cellular senescence | 4 | Otero-Albiol et al.[42] |
| | Cell cycle | 4 | Wang et al.[43] |

Table S2 Results of the Sachs dataset, including the specific number of predicted edges, along with EPR, AUPR, and AUROC evaluation results.

| Methods | #correct | #reverse | #indirect | #unexplain | EPR | AUPR | AUROC | #all |
|---|---|---|---|---|---|---|---|---|
| GES | 4 | 6 | 1 | 9 | 1.111 | 0.318 | 0.132 | 20 |
| LiNGAM | 3 | 7 | 4 | 6 | 0.909 | 0.236 | 0.138 | 20 |
| DirectLiNGAM | 2 | 6 | 3 | 9 | 0.615 | 0.220 | 0.101 | 20 |
| GRNBoost2 | 8 | 5 | 0 | 7 | 2.247 | 0.380 | 0.627 | 20 |
| ANM | 2 | 5 | 5 | 8 | 0.593 | 0.158 | 0.462 | 20 |
| Notears_linear | 8 | 5 | 1 | 6 | 1.000 | 0.611 | 0.358 | 20 |
| Notears_nonlinear | 5 | 6 | 6 | 3 | 1.174 | 0.266 | 0.197 | 20 |
| DAG_GNN | 6 | 4 | 3 | 7 | 1.370 | 0.452 | 0.300 | 20 |
| SCALD | 9 | 5 | 4 | 2 | 2.671 | 0.327 | 0.676 | 20 |

Table S3 The count of nodes within the loops, the total count of all genes, and the count of isolated genes present in the ground truth networks within the BEELINE datasets.

| dataset | #Nodes in loops | #Genes | #Isolated genes |
|---|---|---|---|
| CellType-500 | 26 | 910 | 95 |
| CellType-1000 | 26 | 1410 | 150 |
| NonSpecific-500 | 255 | 910 | 157 |
| NonSpecific-1000 | 255 | 1410 | 272 |
| STRING-500 | 337 | 910 | 399 |
| STRING-1000 | 337 | 1410 | 715 |

**Table S4 Results of ablation studies in six scRNA-seq experiments.** SCALD denotes our comprehensive model, while SCALD_NL represents the non-linear SEM (Structural Equation Modeling) module alone, devoid of any additional constraint components. In other words, SCALD_NL has the ability to predict networks in an arbitrary manner: both stable loops and reinforcing loops can coexist without any imposed constraints

|  |  | EPR | AUPR | AUROC |
|---|---|---|---|---|
| STRING-500 | SCALD | 1.001 | 0.624 | 0.481 |
|  | SCALD_NL | 0.988 | 0.634 | 0.494 |
| STRING-1000 | SCALD | 0.862 | 0.280 | 0.463 |
|  | SCALD_NL | 0.699 | 0.252 | 0.384 |
| CellType-500 | SCALD | 1.233 | 0.339 | 0.597 |
|  | SCALD_NL | 1.119 | 0.292 | 0.522 |
| CellType-1000 | SCALD | 1.299 | 0.322 | 0.575 |
|  | SCALD_NL | 0.935 | 0.293 | 0.511 |
| NonSpecific-500 | SCALD | 1.099 | 0.155 | 0.574 |
|  | SCALD_NL | 1.276 | 0.135 | 0.500 |
| NonSpecific-1000 | SCALD | 1.130 | 0.153 | 0.568 |
|  | SCALD_NL | 1.020 | 0.139 | 0.488 |